\documentclass[12pt]{revtex4}
\usepackage{graphicx}

\begin{document}
\title{Rephasing Invariant Parametrization of Flavor Mixing Matrices}
\author{T. K. Kuo \footnote{ E-mail:
tkkuo@physics.purdue.edu }}
\author{Tae-Hun Lee \footnote{E-mail:
lee109@physics.purdue.edu}}

\affiliation{Physics Department, Purdue University, W. Lafayette, IN
47907}

\begin{abstract}

The three-flavor mixing matrix can be parameterized by the rephasing
invariants $\Gamma_{ijk} = V_{1i} V_{2j} V_{3k}$.  This formulation
brings out the inherent symmetry of the problem and has some
appealing features. Examples illustrating the parametrization and
applications to quark mixing are presented.
\end{abstract}
\maketitle \thispagestyle{empty}

  \pagebreak

\section{Introduction}

It is well-known that the flavor mixing matrices of quarks
($V_{CKM})$ and neutrinos ($V_\nu$) can be multiplied by phase
matrices (rephasing) without changing their physical contents. Thus,
amongst the full set of parameters of these matrices (nine for
$U(3)$ and eight for $SU(3)$), only four are physical.  The choice
of these physical parameters are by no means unique. In fact, a
number of them are in common usage.  One may choose three mixing
angles and a phase, as in the original Cabibbo-Kobayashi-Maskawa
(CKM) parametrization \cite{1}, or the ``standard parametrization"
in the particle data book \cite{2}, or other similar schemes
\cite{3}. For $V_{CKM}$, a very convenient choice turns out to be
the Wolfenstein parametrization \cite{4}, which exhibits the
magnitude of the matrix elements clearly, even though the rephasing
angles are fixed in a specific way.  One could also use the absolute
values $|V_{ij}|$ \cite{5}, which are manifestly rephasing
invariant, although it is not clear which four of these nine should
be singled out. Similarly, another choice is to use four of the nine
rephasing invariants $V_{ik} V_{j \ell} V_{i \ell}^\ast V_{jk}^\ast$
\cite{6}. \vspace{8pt}

In this paper we suggest yet another parametrization based on
rephasing invariants.  Without loss of generality, we consider
only mixing matrices with $\det V = 1$.  There are then six
rephasing invariants $\Gamma_{ijk} = V_{1i} V_{2j} V_{3k}$, ($i,
j, k$) = permutation of $(1, 2, 3)$.  They are shown to satisfy
two simple constraints, leaving us with four independent ones.
These $\Gamma$'s are found to be closely related to the other
rephasing invariants, $|V_{ij}|^2$ and $V_{ik} V_{j\ell} V_{i
\ell}^\ast V_{jk}^\ast$.  However, they retain a lot of the
symmetry inherent in the problem and their construction is equally
valid for $V_{CKM}$ as for $V_\nu$.  These features seem to be
rather appealing, theoretically.  We hope that their use can help
to further our understanding of the flavor mixing problem.
\vspace{8pt}

In Sec.II, we define the rephasing invariants $\Gamma_{ijk}$ and
exhibit the two constraints which reduce the number of independent
parameters to four. Sec.III is devoted to a description of their
detailed properties.  Applications to the quark mixing matrix will
be presented in Sec.IV.  Finally, some concluding remarks are
offered in Sec.V. \pagebreak

\section{ Rephasing Invariant Parametrization}

As we mentioned in the Introduction, there are several known
parametrizations of the flavor mixing matrix.  A common drawback of
these schemes is the lack of uniqueness.  For instance, there are
many ways to choose the mixing angles because of non-commmutativity
\cite{3}.  Similarly, it is not clear which four of the nine
quantities, $|V_{ij}|^2$ or $V_{ik} V_{j \ell} V_{i \ell}^\ast
V_{jk}^\ast$, should be favored.  Despite arguments preferring one
choice over another, it seems fair to say that a general criterion
for a ``best'' set is still absent. We will now introduce yet
another parametrization, which, in our opinion, alleviates the above
problem to a large extent. \vspace{8pt}

We begin by considering, without loss of generality, only mixing matrices which
satisfy
$$ \det V = + 1 ,
\eqno{(1)}
$$

\noindent i.e., only $ SU(3)$, but not $U(3)$, matrices are used.
Note that, while the ``standard parametrization'' satisfies Eq.(1),
the original $KM$ matrix does not. Eq.(1) implies that, in the
rephasing transformation, $V \rightarrow V^\prime = P V P^\prime$,
we can impose on the diagonal phase matrices the conditions, $\det P
= \det P^\prime$ = 1.  It follows immediately that we can construct
a set of six rephasing invariants \cite{7},
$$ \Gamma_{ijk} = V_{1i} V_{2j} V_{3k},
\eqno{(2)}
$$

\noindent where $(i, j, k)$ = permutations of (1, 2, 3).  These
$\Gamma$'s satisfy the constraints ($\det V = 1$),

$$ \sum (\pm) \Gamma_{ijk} = 1 ,
\eqno{(3)}
$$

\noindent where the +(-) sign applies when $(i, j, k)$ is an even
(odd) permutation of (1, 2, 3).  Let us define a matrix $v$,
satisfying

$$ \sum V_{ij} v_{ik} = \sum V_{ji} v_{ki} = \delta_{jk} ,
\eqno{(4)}
$$

\noindent i.e., $v_{ij}$ is the cofactor of $V_{ij}$.  Then, from
$VV^\dag = 1 = \det V$,

$$ V_{ij}^\ast = v_{ij} .
\eqno{(5)}
$$

\noindent E.g., $V_{11}^\ast= V_{22} V_{33} - V_{23} V_{32}$,
$V_{12}^\ast = -(V_{21} V_{33} - V_{23} V_{31})$, etc.  Using
these equalities, we can relate $\Gamma_{ijk}$ to $|V_{\ell
m}|^2$.  For instance,

$$\everymath={\displaystyle}
\begin{array}{ccl}
|V_{12}|^2 & = & V_{12} V_{12}^\ast.  \\
           & = & V_{12} (- V_{21} V_{33} + V_{23} V_{31}) \\
           & = & \Gamma_{231} - \Gamma_{213} \\
\end{array}
\eqno{(6)}
$$

\noindent
Similarly, all the $|V_{ij}|^2$ are equal to the differences of the $\Gamma$'s.
Thus, they must all have the same imaginary part,
$$ \Gamma_{ijk} = R_{ijk} - iJ ,
\eqno{(7)}
$$

\noindent where $R_{ijk}$ is real and $J$ can be identified with the
familiar $CP$-violation measure as follows. We define \cite{6}
\vspace{8pt}

$$\Pi_{\alpha i} = V_{\beta j} V_{\gamma k} V_{\beta k}^\ast V_{\gamma j}^\ast,
\eqno{(8)}
$$

\noindent
where $(\alpha, \beta, \gamma)$ and $(i, j, k)$ are cyclic permutations of
(1, 2, 3), with

$$ Im \Pi_{\alpha i} = J.
\eqno{(9)}
$$
\noindent
Using Eq.(5), we have, for instance,

$$\everymath={\displaystyle}
\begin{array}{ccl}
\Pi_{11} &=& V_{22} V_{33} V_{23}^\ast V_{32}^\ast \\
        &=& (V_{11}^\ast + V_{23} V_{32})(V_{23}^\ast V_{32}^\ast) \\
        &=& \Gamma_{132}^\ast + |V_{23}|^2 |V_{32}|^2 ,
\end{array}
\eqno{(10)}
$$

\noindent establishing $Im \Gamma_{132} = -J$. At the same time,
we may eliminate $V_{23}^\ast V_{32}^\ast$ in $\Pi_{11}$ and find

$$ \Pi_{11} = -\Gamma_{123} + |V_{22}|^2 |V_{33}|^2 ,
\eqno{(11)}
$$

\noindent i.e., the sum ($R_{123} + R_{132}$) is simply related to a
combination of products of the $|V_{ij}|^2$'s. \vspace{8pt}

The above results, with all possible choices of indices, can be collected in a
compact form.  Let us define the matrix.

$$ \everymath={\displaystyle}
W = \left(
\begin{array}{ccc}
|V_{11}|^2  &  |V_{12}|^2    &   |V_{13}|^2 \\

|V_{21}|^2   &  |V_{22}|^2    &  |V_{23}|^2 \\

|V_{31}|^2   &  |V_{32}|^2   &  |V_{33}|^2 \\
\end{array}
\right)
\eqno{(12)}
$$

\noindent together with the matrix $w$ ($w_{ij}$ = cofactor of
$W_{ij}$) defined by

$$ \sum_i W_{ij} w_{ik} = \sum_i W_{ji} w_{ki} = (\det W) \delta_{jk}.
\eqno{(13)}
$$

\noindent
Thus, for instance, $w_{12}$ = - $(|V_{21}|^2 |V_{33}|^2 - |V_{23}|^2 |V_{31}|^2)$.
We further separate the even and odd permutation $R_{ijk}$'s by defining

$$\everymath={\displaystyle}
\begin{array}{cc}
(x_1, x_2, x_3) = (R_{123}, R_{231}, R_{312}) ; \\

(y_1, y_2, y_3) = (R_{132}, R_{213}, R_{321}). \\
\end{array}
\eqno{(14)}
$$

\noindent The relation $\det V = 1$ now reads

$$ (x_1 + x_2 + x_3) - (y_1 + y_2 + y_3) = 1.
\eqno{(15)}
$$

\noindent
The relations between $|V_{ij}|^2$ and $R_{ijk}$ are summarized in

$$\everymath={\displaystyle}
W = \left(
\begin{array}{ccc}
x_1 - y_1  &  x_2 - y_2  &    x_3 - y_3  \\

x_3 - y_2   &   x_1 - y_3   &   x_2 - y_1   \\

x_2 - y_3   &    x_3 - y_1   &    x_1 - y_2   \\
\end{array}
\right),
\eqno{(16)}
$$
\vspace{8pt}

$$\everymath={\displaystyle}
w = \left(
\begin{array}{ccc}
x_1 + y_1   &    x_2 + y_2   &     x_3 + y_3   \\

x_3 + y_2   &    x_1 + y_3   &     x_2 + y_1   \\

x_2 + y_3   &    x_3 + y_1   &    x_1 + y_2    \\
\end{array}
\right).
\eqno{(17)}
$$

\noindent
These equations enable one to switch between the two sets of parameters,
$(x_i, y_j)$ and $|V_{\ell m}|^2$.
\vspace{8pt}

We now turn to the relations between $J$ and $R_{ijk}$

$$ |\Gamma_{ijk}|^2 = |V_{1i}|^2 |V_{2j}|^2 |V_{3k}|^2 = |R_{ijk}|^2 + J^2.
\eqno{(18)}
$$

\noindent Using Eqs.(14, 16), there result six such equations with
$i$ = 1, 2, 3,

$$ (x_i - y_1)(x_i - y_2)(x_i - y_3) - x_i^2 = J^2 ,
\eqno{(19)}
$$

$$ (x_1 - y_i)(x_2 - y_i)(x_3 - y_i) - y_i^2 = J^2.
\eqno{(20)}
$$

\noindent These are consistency conditions which, more
interestingly, may be regarded as cubic equations whose solutions
are the $x$'s and $y$'s.  Thus, $x_i$ are the three solutions of

$$ \xi^3 - (1 + \sum y_j)\xi^2 + (y_1y_2 + y_2y_3 + y_3y_1)\xi - (J^2 + y_1y_2y_3)
= 0 .
\eqno{(21)}
$$

\noindent
Likewise, $y_i$ are those of

$$ \eta^3 + (1 - \sum x_j)\eta^2 + (x_1x_2 + x_2x_3 + x_3x_1)\eta + (J^2 -
x_1x_2x_3) = 0.
\eqno{(22)}
$$

\noindent
It follows that

$$ (x_1 + x_2 + x_3) - (y_1 + y_2 + y_3) = 1 ,
\eqno{(23)}
$$

$$ x_1x_2 + x_2x_3 + x_3x_1 = y_1y_2 + y_2y_3 + y_3y_1 .
\eqno{(24)}
$$

\noindent
In addition,

$$J^2 = x_1x_2x_3 - y_1y_2y_3 .
\eqno{(25)}
$$
Note that Eqs.(24, 25) also follow from Eq.(15) and the identity
$\Gamma_{123}\Gamma_{231}\Gamma_{312}=\Gamma_{132}\Gamma_{213}\Gamma_{321}$.
Thus, a rephasing invariant parametrization of $V$, with $\det V$ =
1, consists of the set $(x_i, y_j)$ subject to the two constraints
in Eqs.(23, 24). Further, the $CP$-violation measure is given by the
very appealing expression in Eq.(25).  One may obtain four
independent parameters out of the set $(x_i, y_j)$ by eliminating
any two of them through Eqs.(23, 24).  However, it is clear that
doing so would lose much of the inherent symmetry of the problem.
\vspace{8pt}

At this juncture it is instructive to compare our results with
those of two flavor mixings.  For $SU(2)$,

$$ V = e^{i \delta \sigma_3} e^{i \theta \sigma_2} e^{i \delta^\prime \sigma_3}
\eqno{(26)}
$$

$$\everymath={\displaystyle}
=\left(
\begin{array}{cc}
g  &  h  \\

-h^\ast  &  g^\ast
\end{array}
\right),
\eqno{(27)}
$$

\noindent $(|g|^2 + |h|^2 = 1)$, and we may parameterize $V$
either by $\theta$ or by one of the $|V_{ij}|$'s, say, $|V_{11}|^2
= |g|^2 = \cos^2 \theta$.  However, one may equally well have used
the (real) rephasing invariant parameters defined by

$$ x = \Gamma_{12} = V_{11} V_{22} = \cos^2 \theta ,
\eqno{(28)}
$$

$$ y = \Gamma_{21} = V_{12} V_{21} = - \sin^2 \theta,
\eqno{(29)}
$$

$$ x - y = \det V = 1 .
\eqno{(30)}
$$

\noindent
While the generalizations to three flavors of the first two parameterizations
are well-known, that of the third leads to the set $(x_i, y_j)$ which was
studied above.

\pagebreak

\section{Properties of the Parametrization}

We now turn to some detailed properties of the parameters $(x_i,
y_j)$.  Let's start by establishing the range of their values.
First, from Eq.(16), we have

$$ (y_1, y_2, y_3) \leq (x_1, x_2, x_3) .
\eqno{(31)}
$$

\noindent Next, with $W_{ij} \leq 1$ and $|w_{ij}| \leq 1$, and
relations Eqs.(16, 17) such as $x_1 = \frac{1}{2} (W_{11} +
w_{11})$, etc., we readily find

$$ -1 \leq (x_i, y_j) \leq 1 .
\eqno{(32)}
$$

Consistency of the constraint, Eq.(24) with Eqs.(31, 32) can now
be used to establish that at most one $x_i$ can be negative (and,
similarly, only one $y_j$ can be positive) and that

$$ x_1x_2 + x_2x_3 + x_3x_1\geq 0 .
\eqno{(33)}
$$

\noindent
Finally, it is not hard to deduce that

$$ 0 \leq (x_1 + x_2 + x_3) \leq 1 ,
\eqno{(34)}
$$

$$ -1 \leq (y_1 + y_2 + y_3) \leq 0 .
\eqno{(35)}
$$

To summarize, the parameters $(x_i, y_j)$ are distributed in the
interval [-1, 1], with $x_i \geq y_j$, for all $(i, j)$.  Also, $0
\leq \sum x_i \leq 1$, with at most one negative $x_i$, while $-1
\leq \sum y_j \leq 0$, with at most one positive $y_j$.
\vspace{8pt}

Turning to the matrices $W$ and $w$, with $Ww^T = (\det W) I$, we
find

$$ \det W = (x_1^2 + x_2^2 + x_3^2) - (y_1^2 + y_2^2 + y_3^2),
\eqno{(36)}
$$

\noindent which, by Eqs.(23, 24), reduces to

$$ \det W = (x_1 + x_2 + x_3) + (y_1 + y_2 + y_3) .
\eqno{(37)}
$$

\noindent It follows, by Eqs.(34, 35), that

$$ -1 \leq \det W \leq 1 .
\eqno{(38)}
$$

\noindent It is interesting to note that, while the elements of
any row or column in $W$ sum up to unity, the corresponding sum in
$w$ is equal to $\det W$, $\sum_i w_{iI} = \sum_i w_{Ii} = \det
W$.
\vspace{8pt}

The properties discussed above also suggest an interesting
relation between three flavor and two flavor mixing, as contained
in the correspondence: $(\sum x_i) \rightarrow x$ and $(\sum y_j)
\rightarrow y$.  For two flavors, $\det W $ $(= x + y = \cos 2
\theta$) can be regarded as a measure of the deviation of the
mixing from identity, with $\det W = 0$ at maximal mixing.  It
seems reasonable to use $\det W$ as a measure of the total amount
of mixing for three flavors, with a necessary condition for
maximal mixing again being $\det W = 0$. \vspace{8pt}

We now consider two concrete examples which should help to illuminate the
nature of the $(x_i, y_j)$ parametrization.
\vspace{12pt}

\noindent
A)  $x_1 = 1, x_2 = x_3 = y_1$ = ... = 0.
\vspace{5pt}

This solution of course corresponds to the identity matrix, $V = I
$, with $ \det W $ = +1.  Cyclic permutations of the states
generate equivalent solutions, with some $x_i = 1$ while all other
parameters vanish.  An exchange of the states switches the roles
of $x$ with $y$, resulting in solutions such as $y_1 = -1, x_1 =
... = 0$, with $\det W = -1$. \vspace{8pt}

The physical quark mixing matrix, $V_{CKM}$, is very close to the identity
matrix, with $x_1 \simeq 1$ and all other parameters $\simeq 0$.  We will give
a detailed description of $V_{CKM}$ in the next section.
\vspace{8pt}

\noindent
B) $x_1 = x_2 = x_3 = 1/6, y_1 = y_2 = y_3 = -1/6.$
\vspace{8pt}

The solution exhibits maximal symmetry for three flavor mixing, with

$$ \everymath={\displaystyle}
W = \left(
\begin{array}{ccc}
1/3   &    1/3  &    1/3  \\

1/3   &    1/3   &    1/3  \\

1/3   &   1/3   &   1/3  \\
\end{array}
\right)
\eqno{(39)}
$$

\noindent $\det W$ = 0 and $w_{ij} = 0$.  It is the three flavor
generalization of the maximal mixing solution for the flavors,
where the 2 $\times$ 2 $W $matrix is given by

$$\everymath={\displaystyle}
W = \left(
\begin{array}{cc}
1/2  &    1/2  \\

1/2   &   1/2  \\
\end{array}
\right)
\eqno({40})
$$

\noindent so that $\theta =\pi/4$ and $\det W$ = 0. \vspace{8pt}

The mixing matrix corresponding to the maximal symmetry solution,
Eq.(39), can also be written down, provided one chooses a specific
phase.  If we use the ``standard'' parametrization, then

$$\everymath={\displaystyle}
V = \left(
\begin{array}{ccc}
\frac{1}{\sqrt{3}}  &  \frac{1}{\sqrt{3}}  &  -i\frac{1}{\sqrt{3}}  \\

\frac{-e^{i \alpha}}{\sqrt{3}}  &  \frac{e^{-i \alpha}}{\sqrt{3}}  &  \frac{1}{\sqrt{3}}   \\

\frac{e^{-i \alpha}}{\sqrt{3}}  &  -\frac{e^{i \alpha}}{\sqrt{3}}
&  \frac{1}{\sqrt{3}
} \\
\end{array}
\right),
~~~\alpha = \pi/6.
\eqno{(41)}
$$

\noindent This solution is now known as being
``trimaximal''\cite{8}. It is a particular case of a bimaximal
solution \cite{9}, with $\theta_{12}=\theta_{23}=\pi/4$, but
$\sin\theta_{13}=1/\sqrt{3}$ and $\delta=\pi/2$. It is noteworthy
that the matrix $V$ is complex. Indeed, this solution was known
\cite{6} to give rise to the maximally allowed value for $J^2$,
given by Eq.(25)

$$ J^2 = 1/108 ,
\eqno{(42)}
$$

\noindent Within the present parametrization, we may demonstrate
this fact by considering the variation of $J^2 = x_1x_2x_3 -
y_1y_2y_3$, for arbitrary $\delta x_i$ and $\delta y_j$ but subject
to the constraint $\sum (\delta x_i) - \sum(\delta y_j)$ = 0.  At a
symmetric point, $x_i= -y_j$, for all $ (i,j) $, so that also $x_i
x_j = y_\ell y_m$,

$$\everymath={\displaystyle}
\begin{array}{rcl}
$$ \delta J^2 &=& (\delta x_1 + \delta x_2 + \delta x_3) x_i x_j - (\delta y_1 + \delta y_2 + \delta y_3) y_\ell y_m \\
              &=& 0 .
\end{array}
\eqno{(43)}
$$

Physically, the neutrino mixing matrix is close to being bimaximal but with a
small $\theta_{13}$, and with $\delta$ completely unknown.  It is tempting to
speculate that there is a common origin (renormalization being a prime candidate)
of the deviations of $\theta_{12}, \theta_{13}$ and $\delta$ from the maximal
symmetry solution.  If this is correct, then, from the known physical values of
$\theta_{12}$ and $\theta_{13}$, we would have a means to calculate the phase
$\delta$.
\vspace{8pt}

\section{Applications to $V_{CKM}$}

For the CKM matrix, a particularly useful (approximate)
parametrization is due to Wolfenstein \cite{4}, with $\lambda\simeq$
0.22,

$$\everymath={\displaystyle}
V_{CKM} = \left(
\begin{array}{ccc}
1 - \lambda^2/2   &       \lambda     &      A \lambda^3 (\rho - i\eta) \\

   -\lambda       &    1 - \lambda^2/2     &        A \lambda^2 \\

A \lambda^3 (1 - \rho - i \eta)    &        -A \lambda^2       &        1     \\
\end{array}
\right)
 + O(\lambda^4)
\eqno{(44)}
$$

\noindent More accurate formulas are also available \cite{10}.  The
matrix is simple in form yet it captures all of the essence of the
quark mixing. Note, however, $\det V \neq 1$. \vspace{8pt}

To arrive at Eq.(44), one needs to choose the phases so that only
$V_{13}$ and $V_{31}$ are complex.  We note that a rephasing
invariant parametrization can be constructed in terms of the $ W $
matrix.

$$\everymath={\displaystyle}
W_{CKM} = \left(
\begin{array}{ccc}
1 - a^2 - b^2      &          a^2         &            b^2 \\

a^2 - e^2      &     1 -   a^2 - d^2 + e^2     &       d^2  \\

b^2 + e^2       &          d^2 - e^2      &        1 - b^2 - d^2   \\
\end{array}
\right).
\eqno{(45)}
$$

\noindent
In this construction, we have incorporated the unitarity conditions which also
implies the relations $W_{12} - W_{21} = -(W_{13} - W_{31}) = W_{23} - W_{32}$.
The choice $W_{31} - W_{13} = e^2 \geq 0$ is made here in accordance with the
experimental $|V_{CKM}|$ values.  To make connections to $V_{CKM}$ and to
exhibit the order of magnitudes of the various parameters, we may write

$$\everymath={\displaystyle}
\begin{array}{llll}
a^2   =  \lambda^2 , \\

b^2   =   B^2 \lambda^6 ,   \\

d^2   =   D^2 \lambda^4 ,  \\

e^2   =   E^2 \lambda^6 .  \\
\end{array}
\eqno{(46)}
$$

\noindent These relations define $(\lambda^2, B^2, D^2, E^2)$,
with $(B^2, D^2, E^2)$ all being of order unity.  We emphasize
that Eq.(45), with the values given in Eq.(46), is an exact
parametrization, and not an expansion in $\lambda$.  Thus, Eq.(46)
may be regarded as a mnemonic device to remind us of the physical
values of the parameters $(a^2, b^2, d^2, e^2)$.  At the same
time, once we have a quantity expressed in terms of them, it may
also be used to obtain an expansion in $\lambda$. Thus,
Eqs.(45,46) is a rephasing invariant generalization of the
Wolfenstein parametrization, whereby higher order terms in
$\lambda$ can be read off directly. \vspace{8pt}

From Eq.(45), we can calculate the elements of $ w $, and hence the
parameters $(x_i, y_j)$.  We have

$$\everymath={\displaystyle}
\begin{array}{llllll}
2x_1 = 2 - 2(b^2 + a^2 + d^2) + e^2 + \bar{w}  \\

2x_2 =  e^2 + \bar{w}  \\

2x_3 = -e^2 + \bar{w}    \\

2y_1 = -2d^2 + e^2 + \bar{w}  \\

2y_2 = -2a^2 + e^2 + \bar{w}  \\

2y_3 = -2b^2 - e^2 + \bar{w}   \\
\end{array}
\eqno{(47)}
$$

\noindent
where $\bar{w}$ is defined by

$$ \bar{w} = a^2d^2 + b^2(a^2 + d^2 - e^2).
\eqno{(48)}
$$

\noindent Eqs.(47, 48) are exact.  But it is useful to get an
order of magnitude estimate by putting in the values of Eq.(46),
we see that $x_1 \simeq O(1)$, $(x_2, x_3) \simeq O(\lambda^6)$,
$y_1 \simeq O(\lambda^4)$, $y_2 \simeq O(\lambda^2)$, $y_3 \simeq
O(\lambda^6).$  $\bar{w}$ contains terms up to $O(\lambda^{12})$,
with the leading order term being $a^2d^2$.  The constraint $\sum
x_i - \sum y_j = 1$ is easily verified.  The constraints,
Eqs.(23,24), which are valid to all orders in $\lambda$, lead to
simple approximate relations (with $O(\lambda^2)$ corrections)
$$ x_1 \simeq 1 + y_1 + y_2 ,
\eqno{(49)}
$$

$$ x_2 + x_3 \simeq y_1 y_2 .
\eqno{(50)}
$$
\vspace{8pt} In terms of the parameterizations in Eq.(45), we can
readily find the $CP$-violation measure
$$J^2=\frac{1}{4}[-a^4d^4-b^4-(b^2+e^2)^2+2b^2(b^2+e^2)+2a^2d^2(2b^2+e^2)]- \Delta
\eqno{(51)}
$$
$$\Delta=\frac{1}{4}(\bar{w}-a^2d^2)[(\bar{w}+a^2d^2)+2e^2]
\eqno{(52)}
$$
\noindent where $\bar{w}$ is defined in Eq.(48). In Eq.(51),
$\Delta$ is $O(\lambda^2)$ compared to the term in the square
bracket, which can be shown to be the $4\times (area)^2$ of a
triangle with sides $(ad,b,\sqrt{b^2+e^2})$. This result is of
course well-known in connection with the study of the unitarity
triangles, which we will discuss in the following. \vspace{12pt}

Consider the unitarity conditions, $\sum V_{ij} V_{ik}^\ast =
\delta_{jk}$.  Within the context of the present discussion, for $j = k$, they are
rephasing invariant and, with Eq.(5), reduce to $\sum (\pm) \Gamma_{ijk} = 1$,
while for $j \not= k$, the conditions are rephasing dependent, but are
identities if Eq.(5) is used.  Thus, tests of the unitarity triangles
amounts to those of Eq.(5).  It is well-known that the most interesting
relation is

$$ V_{ud} V_{ub}^\ast + V_{cd} V_{cb}^\ast + V_{td} V_{tb}^\ast = 0 ,
\eqno{(53)}
$$

\noindent
or,

$$ V_{11} V_{13}^\ast + V_{21} V_{23}^\ast + V_{31} V_{33}^\ast = 0 .
\eqno{(54)}
$$

\noindent We can turn this equation into one with only rephasing
invariants by multiplying, for instance, by $V_{21}^\ast V_{23}$:

$$ V_{11} V_{13}^\ast V_{21}^\ast V_{23} + |V_{21}|^2 |V_{23}|^2 + V_{31} V_{33}^\ast
V_{21}^\ast V_{23} = 0
\eqno{(55)}
$$

\noindent
This relation is displayed in Fig. 1.  It is the usual unitarity triangle rotated
and magnified (by $|V_{21}| |V_{23}|)$.  It has a base $|V_{21}|^2 |V_{23}|^2$.
The other two sides are given by

$$\everymath={\displaystyle}
\begin{array}{lll}
V_{11} V_{13}^\ast V_{21}^\ast V_{23} &=& -\Gamma_{312}^\ast +
     |V_{13}^2| |V_{21}|^2   \\
                                      & \cong & - x_3 - iJ ,
\end{array}
\eqno{(56)}
$$

$$\everymath={\displaystyle}
\begin{array}{lll}
V_{31} V_{33}^\ast V_{21}^\ast V_{23} & = & - \Gamma_{231} + |V_{23}|^2 |V_{31}|^2 \\
                                     & \cong &  -x_2 + iJ  ,  \\
\end{array}
\eqno{(57)}
$$

\noindent where we have used
$|V_{13}^2|V_{21}|^2=(x_3-y_3)(x_3-y_2)\ll x_3$, and
$|V_{23}^2|V_{31}|^2=(x_2-y_1)(x_2-y_3)\ll x_2$. Thus, the
triangle in Fig. 1 has height $J$, with the lengths of the two
sides being approximately $(x_3^2 + J^2)^{1/2}$ and $(x_2^2 +
J^2)^{1/2}$. Also, the base line has length $\cong x_2 + x_3$,
according to Eq.(50).  It follows that

$$ \tan \beta \cong J/x_2 ,
\eqno{(58)}
$$

$$ \tan \gamma \cong J/x_3 .
\eqno{(59)}
$$

\noindent
A similar construction (by choosing a different real base line) yields

$$ \tan \alpha \cong J/y_3 .
\eqno{(60)}
$$
\vspace{0pt} \noindent In other words, the angles
$(\alpha,\beta,\gamma)$ are simply the (approximate) phase angles of
the rephasing invariants $\Gamma_{231}^\ast, \Gamma_{312}^\ast$ and
$\Gamma_{321}^\ast$, respectively.

\begin{figure}
\includegraphics[width=5.2 in, height=3.6 in]{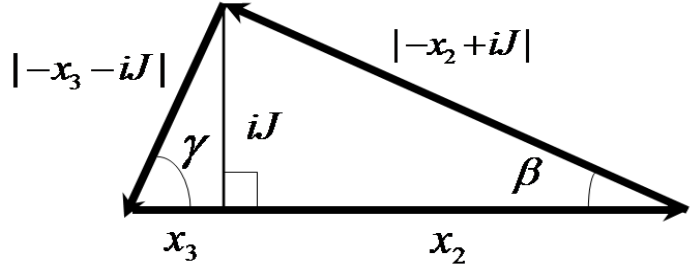}
\caption{Magnified unitary triangle with sides
$|V_{21}|^2|V_{23}|^2$, $V_{31}V_{33}^\ast V_{21}^\ast V_{23}$ and
$V_{11}V_{13}^\ast V_{21}^\ast V_{23}$. Their approximate lengths
are as labelled.}
\end{figure}

Experimentally, $CP$-violating processes seem to indicate that
$\alpha\cong\pi/2$ \cite{11}. This is a very intriguing result since
it implies that $y_3 $ is much smaller than $O(\lambda^6)$, the
expected ``natural'' value. To the extent that all of the above
results are valid to $O(\lambda^2)$, we take $y_3 = O(\lambda^8)$.
From Eq.(47), $\alpha \cong \pi /2$ implies that
$$ a^2d^2\cong 2b^2 + e^2
\eqno{(61)} $$
\noindent
or,
$$ |V_{us}|^2|V_{cb}|^2 \cong |V_{ub}|^2 + |V_{td}|^2
\eqno{(62)}
$$
\noindent Also, $y_3 \cong 0$ means that $x_3 \cong b^2$, $x_2
\cong b^2 + e^2$, from Eqs.(16,45). Thus, a particularly simple
set of parameters result,
$$\everymath={\displaystyle}
\begin{array}{cc}
x_1 \cong 1-|V_{us}|^2-|V_{cb}|^2, \\
x_2 \cong  |V_{td}|^2, \\
x_3 \cong  |V_{ub}|^2, \\
y_1 \cong - |V_{cb}|^2, \\
y_2 \cong - |V_{us}|^2, \\
y_3 \cong  0, \\
\end{array}
\eqno{(63)}
$$
\noindent assuming $\alpha \cong \pi/2$. All above relations are
accurate to $O(\lambda^2)$. In addition, for $\alpha = \pi/2$,
$\tan\beta \tan\gamma = 1$, so that from Eqs.(58,59) (or from
Eq.(25)) we find
$$
J^2 \cong x_2x_3 \cong |V_{td}|^2|V_{ub}|^2.
\eqno{(64)}
$$
\noindent The above relations reveal that for $V_{CKM}$, the
parameters $(x_i, y_j)$ have particularly simple relations with
the directly measured quantities $|V_{ij}|^2$ and $(\alpha, \beta,
\gamma)$. Whether there is a deeper meaning behind the pattern in
Eq.(63) remains to be seen. \vspace{8pt}

\pagebreak

\section{ Conclusion}

In this paper, we propose to parameterize a three flavor mixing
matrix by $\Gamma_{ijk}$ (Eq. (2)), which are rephasing invariant
when we demand, without loss of generality, that $\det V =  1$.
All of the $\Gamma$'s have the same imaginary part, $-J$, which is
the $CP$-violation measure.  The six real parts of $\Gamma$
satisfy two constraints (Eqs.(23, 24)), resulting in four
independent ones, as expected.  In addition, $J^2$ is given in a
very symmetric expression, Eq. (25). \vspace{8pt}

The $\Gamma$-parametrization is characterized by its symmetry,
which is a reflection of the inherent property of the three-flavor
mixing.  With its help we are able to identify a mixing pattern of
``maximal symmetry'', in Eqs.(39, 41). Its resemblance to the
neutrino mixing matrix seems to suggest a possible origin of the
latter.  This possibility will be explored. \vspace{8pt}

The relation between the $(x_i,y_j)$ parameters and $|V_{ij}|^2$
was discussed in detail.  As an application we find explicit
$(x_i,y_j)$ values corresponding to the physical $V_{CKM}$. It is
shown that all of the measurable quantities $(|V_{ij}|^2$, phase
angle$(\alpha, \beta, \gamma)$) are directly related to the $(x_i,
y_j)$ variables. Of the three $x$-values, one is close to unity
and the other two are small $(O(\lambda^6))$, while the three
$y$-values are of order $O(\lambda^2), O(\lambda^4)$ and
$O(\lambda^8)$, respectively. To a good approximation, $\alpha
\cong \pi/2$, it is found that $(x_2, x_3, y_1, y_2)$ are simply
equal to $(|V_{td}|^2, |V_{ub}|^2, -|V_{cb}|^2, -|V_{us}|^2)$.
\vspace{8pt}

The use of rephasing invariants should be useful in other
problems, for instance, in parameterizing the mass matrices.  We
hope to return to this topic in the future. \vspace{24pt}

This work is supported in part by DOE grant NO. DE-FG02-91ER40681.

\pagebreak


\begin{thebibliography}{10}
\bibitem{1}
M. Kobayashi, and T. Maskawa, Prog. Th. Phys. {\bf 49}, 652 (1973).
\bibitem{2} Particle Data Group, Phys. Lett. {\bf B952}, 130 (2004); L. L.
Chau and W. Y. Keung, Phys. Rev. Lett. {\bf 53}, 1802 (1984); H.
Harari and M. Leurer, Phys. Lett. {\bf B181}, 123 (1986).
\bibitem{3} See, e.g., H. Fritzsch and Zhi-Zhong Xing, Prog. Part.
Nucl. Phys. {\bf 45}, 1 (2000). \vspace{5pt}
\bibitem{4} L. Wolfenstein, Phys. Rev. Lett. {\bf 51}, 1945 (1983).
\bibitem{5} C. Hamzaoui, Phys. Rev. Lett {\bf 61}, 35 (1988); G. C.
Branco and L. Lavoura, Phys. Lett. {\bf B208}, 123 (1988).
\bibitem{6} C. Jarlskog, Phys. Rev. Lett. {\bf 55}, 1039(1985);
also in $CP$ Violation, Edited by C. Jarlskog (World Scientific,
Singapore, 1989).
\bibitem{7} S. Chang and T. K. Kuo, Phys. Rev. {\bf D66}, 111302(R), (2002).
\bibitem{8} V. Barger, K. Whisnant and R. J. N. Phillips, Phys. Rev. {\bf D24}, 538 (1981); C. Giunti, C. W. Kim and J. D. Kim, Phys. Lett. {\bf B352}, 357 (1995); P. F. Harrison, D. H. Perkins, W. G. Scott, Phys. Lett. {\bf B349}, 137 (1995).
\bibitem{9} F. Vissani, hep-ph/9708483; V. Barger, S. Pakvasa, T. J.
Weiler and K. Whisnant, Phys. Lett. {\bf B437}, 107 (1998); A. J.
Baltz, A. S. Goldhaber and M. Goldhaber, Phys. Rev. Lett. {\bf 81},
5730 (1998)
\bibitem{10} A. J. Buras, M. E. Lautenbacher and G.
Ostermaier, Phys. Rev. {\bf D50}, 3433 (1994).
\bibitem{11} See, e.g., Z. Ligeti, hep-ph/0408267.
\end{thebibliography}
\end{document}